\def\be{\begin{equation}}
\def\ee{\end{equation}}
\def\bea{\begin{eqnarray}}
\def\eea{\end{eqnarray}}
\def\beg{\begin{align}}
\def\eeg{\end{align}}

\documentclass[a4paper,11pt]{article}
\pdfoutput=1 

\usepackage{jheppub} 

\usepackage[T1]{fontenc} 

\title{\boldmath Efimov-like Resonances in Planar QED}

\author[a,1]{K. Abhinav\note{Corresponding author.}}
\author[b]{P. K. Panigrahi}

\affiliation[a]{S N Bose National Centre for Basic Sciences, JD Block, Sector III, Salt Lake, Kolkata 700106,  India}
\affiliation[b]{Indian Institute of Science Education And Research Kolkata, Mohanpur-741246, West Bengal, India}

\emailAdd{kumar.abhinav@bose.ac.in}
\emailAdd{pprasanta@iiserkol.ac.in}

\abstract{It is shown that planar topological effective gauge theory with dynamics, acquires corrections to angular
momentum beyond the well-known topological photon spin, the latter arising from interactions with parity-breaking
massive fermions. In the non-relativistic limit, a first quantized Schr\"odinger representation is possible where 
the topological and kinetic terms decouple, the latter contributing to angular momentum and compete with the centrifugal
barrier. This results in shallow resonances of Efimov kind, which may be verified in planar physical systems.}

\begin{document} 
\maketitle
\flushbottom

\section{Introduction}
Planar (2+1 dimensional) quantum electrodynamics (QED) displays enhanced renormalizability
\cite{Ren}, tractable infrared behavior \cite{Inf} and inherent topology \cite{Red,TopM1,TopM2,Boy}.
The last property makes the theory relevant to phenomena as fractional Hall effect \cite{QHE1,QHE2},
topological insulators \cite{TI1,TI2}, anyon superconductivity \cite{Any1,Any2} and many
more. The corresponding physics is effectively captured by the induced Chern-Simons (CS) term,

\be
\mathcal{L}_{CS}:=\frac{\mu}{2}\epsilon^{\mu\nu\rho}a_{\mu}\partial_{\nu}a_{\rho},\nonumber
\ee
with gauge field $a_\mu$, imbibing the planar photon with spin $\mu/\vert \mu\vert$ \cite{TopM2,Boy},
though it does not effect dynamics of the gauge field \cite{Dunne}. Such an effective theory supports
excitonic states \cite{H} in presence of quantum fluctuations, which can survive the additional
gauge dynamics \cite{Own1,Own2}, even at finite temperature \cite{Own2}.   
\paragraph*{}The fact that planar fermions can have non-trivial properties in presence of gauge interaction
have been realized extensively \cite{H,H2,Red,TopM1,TopM2}, which can be expressed as an effective theory in the gauge
sector \cite{H,Own2}. The latter approach is essentially a linear response treatment, characterized by
the inherent topological aspects of the system, protected beyond first order quantum corrections
by the celebrated Coleman-Hill theorem \cite{CH}.
\paragraph*{}An effective gauge theory, capturing linear response of fermions, collectively
represents the dynamics of the latter species. The CS term carries the signature of the
parity-breaking fermion mass $m$, in addition to the usual transverse vacuum polarization contribution
$\propto F_{\mu\nu}F^{\mu\nu}$ adding to the dynamics of the gauge field. A gauge theory, entirely
defined by the CS term, was shown to satisfy the Schr\"odinger symmetry \cite{H}, and thereby
is separable into spatial and temporal sectors independently \cite{H2}. When pure CS
term is present at the {\it tree level}, the inherent topology imbibed into the fermion mass, together give
rise to the parity anomaly \cite{H}, expressed as a correction to the fermion angular momentum
\cite{H,H2}. 
\paragraph*{}The effect on gauge angular momentum, which is gauge dependent \cite{Greiner}, due
to coupling with parity-breaking massive fermions in a plane, has not been evaluated yet. However, at the
level of effective theories, which become important in case of linear response studies of properties like
Hall effect \cite{QH}, conductance \cite{TI} and spin control \cite{Own1}, such an analysis become important. Moreover, 
study of gravitational models in planar systems \cite{Own3,Polyakov}, wherein gravity is equivalent to CS gauge 
fields \cite{Witten}, symmetries of the latter are of deep interest. More importantly, the inherent 
Lorentz symmetry of this system includes the CS contribution, emerging through interaction. 
The corresponding physical analogues, like graphene and topological insulators (TIs), are low-energy
systems \cite{Neto,TI}, validating a linear response treatment.
\paragraph*{}In the following, the effect of quantum correction to the orbital angular momentum (OAM) of gauge field will be
evaluated in Section 2. The corresponding topological influence will be marked
out, the latter contributing to the gauge spin. In a non-relativistic approximation
in suitable gauge in Section 3, will lead to a `Schr\"odinger equation of photon', with {\it local} shift to 
the centrifugal potential, leading to shallow resonances with Efimov signature. This owes solely to the tree-level dynamics,
and is devoid of topological effects or quantum fluctuations, complementing Coulomb interaction. Then we conclude in
Section 4, suggesting possible physical realizations of such theories.

\section{Corrections to angular momentum}
Here we consider an effective gauge theory, with both dynamics and CS topology at the
tree level, along with a vacuum polarization term representing the linear response of the interactions.
The corresponding Lagrangian has the form,

\be
\mathcal{L}\equiv-\frac{1}{4}F_{\mu\nu}F^{\mu\nu}+\frac{\mu}{2}\epsilon^{\mu\nu\rho}a_{\mu}\partial_{\nu}a_{\rho}+\frac{1}{2}a_{\mu}\Pi^{\mu\nu}a_{\nu},\label{L2}
\ee   
where, the momentum space representation of the vacuum polarization tensor is \cite{Rao},

\bea
\Pi^{\mu\nu}(q)&=&\Pi_e(q)\left(q^\mu q^\nu-g^{\mu\nu}q^2\right)+\Pi_o(q)\epsilon^{\mu\nu\rho}q_{\rho},\nonumber\\
\Pi_e(q)&=&\frac{e^2}{4\pi}\left[\frac{1}{|q|}\left(\frac{1}{4}+\frac{m^2}{q^2}\right)\log\left(\frac{2\vert m\vert+\vert q\vert}{2\vert m\vert-\vert q\vert}\right)-\frac{|m|}{q^2}\right],\nonumber\\
\Pi_o(q)&=&-i\frac{m}{4\pi}\frac{e^2}{|q|}\log\left(\frac{2\vert m\vert+\vert q\vert}{2\vert m\vert-\vert q\vert}\right).\label{04}
\eea
The corresponding energy-momentum tensor can be evaluated, generically, as the canonical
conjugate to the metric tensor $g^{\mu\nu}$, from the following variation of the Lagrangian,

\be
T^{\mu\nu}=\frac{2}{\sqrt{-g}}\frac{\delta}{\delta g_{\mu\nu}}\left(\sqrt{-g}\mathcal{L}\right)\equiv2\frac{\delta \mathcal{L}}{\delta g_{\mu\nu}}+g^{\mu\nu}\mathcal{L},\label{L2a}
\ee
which is symmetric by construction \cite{IZ}. For the present Lagrangian, the energy-momentum 
tensor has the form,

\begin{align}
T^{\mu\nu}\equiv&-F^{\mu\alpha}F^{\nu}_{~~\alpha}+\mu\left[\epsilon^{\mu\alpha\beta}a^{\nu}\partial_{\alpha}a_{\beta}+\epsilon^{\alpha\mu\beta}a_{\alpha}\partial^{\nu}a_{\beta}+\epsilon^{\alpha\beta\mu}a_{\alpha}\partial_{\beta}a^{\nu}+(\mu\leftrightarrow\nu)\right]+2\Big[a^{\mu}\Pi^{\nu\alpha}a_{\alpha}\nonumber\\
&+(\mu\leftrightarrow\nu)\Big]+2a_{\alpha}V^{\alpha\beta}\partial^{\mu}\frac{1}{\square}\partial^{\nu}a_{\beta}2g^{\mu\nu}\mathcal{L}-2\xi(\partial.a)\left[\partial^{\mu}a^{\nu}+(\mu\leftrightarrow\nu)\right];\label{L3}
\end{align}
with,

\begin{align}
V^{\alpha\beta}(q)&=P^{\alpha\beta}\left[\frac{\vert m\vert}{\pi\left(4m^2-q^2\right)}\left(\frac{1}{4}+\frac{m^2}{q^2}\right)+i\frac{2m}{q^2}\Pi_o(q)-\Pi_e(q)\right]\nonumber\\
&~~~~+\epsilon^{\alpha\rho\beta}q_{\rho}\left[i\frac{m\vert m\vert}{\pi\left(4m^2-q^2\right)}-\Pi_o(q)\right];\label{L4}\\
{\rm where},&\nonumber\\
P^{\mu\nu}(q)&=q^{\mu}q^{\nu}-\eta^{\mu\nu}q^2,\nonumber
\end{align}
in the momentum representation, in the covariant $R_\xi$ gauge. At low energies, the
form factors $\Pi_{e,o}$ can be considered as finite constants \cite{Inf,TopM1,TopM2,Red,Boy,H}.
\paragraph*{}The OAM tensor for any field is defined as \cite{IZ},

\be
{\cal M}^{\mu\nu\rho}:=\int d^2x\left(T^{\mu\rho}x^\nu-T^{\mu\nu}x^\rho\right),\nonumber
\ee
leading to the OAM pseudo-vector in planar systems as, 

\be
J=\int d^2x{\bf x}\times{\bf T},~~~T^k=T^{0k},\label{L1}
\ee
with spatial indexes expressed in Latin. From Eq. \ref{L3}, `spatial' components of the
energy-momentum tensor can be written as,

\begin{align}
T^{0k}\equiv&-F^{0i}F^{k}_{~~i}+\mu\Big[\epsilon^{0ij}a^{k}\partial_{i}a_{j}+\epsilon^{i0j}a_{i}\partial^{k}a_{j}+\epsilon^{ij0}a_{i}\partial_{j}a^{k}+\epsilon^{k\alpha\beta}a^0\partial_{\alpha}a_{\beta}\nonumber\\
&+\epsilon^{\alpha k\beta}a_{\alpha}\partial^0a_{\beta}+\epsilon^{\alpha\beta k}a_{\alpha}\partial_{\beta}a^0\Big]+2a^0\Pi^{k\alpha}a_{\alpha}+2a^k\Pi^{0\alpha}a_{\alpha}\nonumber\\
&+2a_{\alpha}V^{\alpha\beta}\partial^{0}\frac{1}{\square}\partial^{k}a_{\beta}-2\xi(\partial.a)\left(\partial^{0}a^{k}+\partial^{k}a^{0}\right).\label{L5}
\end{align}
which finally leads to the evaluation of OAM of spatial component of the gauge field 
through the evaluation of the commutator $\left[J,a^i(x)\right]$ \cite{H2}, which can be
resolved into commutators of gauge field and its derivatives. The inherent gauge redundancy
of the system makes the fundamental commutators dependent on the choice of the gauge. For the
future convenience, we choose the `physical' Coulomb gauge $\nabla\cdot{\bf a}=0$, implementing
the {\it spatial} transverse nature of the gauge field, leading to \cite{AD1}, 

\bea
\left[\dot{a}^i(x),a^j(y)\right]_{x_0=y_0}&=&i\delta^{ij}_{TR}({\bf x}-{\bf y}); \quad{\rm where},\label{04}\\
\delta^{ij}_{TR}({\bf x}-{\bf y})&=&\delta^{ij}\delta\left({\bf x}-{\bf y}\right)+\partial^i\frac{1}{\boldsymbol{\nabla}^2}\partial^j\delta\left({\bf x}-{\bf y}\right),\nonumber
\eea
with rest of the combinations vanishing. The choice of a {\it non-relativistic} gauge is justified even
though we have been explicitly employing the covariant $R_{\xi}$ gauge till now, as we are dealing
with the OAM, which is spatial. Equivallently, the particular covariant gauge choice can be considered to have a
spatial subsector of the desired form.  Finally, the OAM of the spatial gauge field component turns out to
be,

\bea
\left[J,a^{k}\right]\equiv&-&L\left[\eta^{\alpha k}-i\frac{1}{q^2}V^{\alpha k}\right]a_{\alpha}-i\left\{1-i\Pi_e\right\}\left[\epsilon^{kj}a^j-\partial^k\left\{{\bf x}\times{\bf a}\right\}\right]\nonumber\\
&-&i\left\{\mu-i\Pi_o\right\}x^ka_0+i\Pi_e\epsilon^{kj}x^j\dot{a}_0+(\text{Higher derivative terms}).\label{L6}
\eea
There are a few points to be noted here. The naiv\'e tree-level OAM
$L:=-i{\bf x}\times\boldsymbol{\nabla}_x$ gets modified by a term proportional to $V^{\mu\nu}$, 
the latter incorporating quantum corrections through $\Pi_{e,o}$, along-with
contribution from tree-level dynamics. Both form factors are singular at the two-fermion threshold $q^2=4m^2$, as
signature of the planar exciton \cite{H,Own2}. The next two terms represent anomalous contribution
to the planar OAM. The second is the planar analogue 
of photon spin in 3+1 dimensions \cite{Greiner}, whereas the third represents the topological spin
of planar gauge fields, with both containing respective quantum corrections. The latter 
also corresponds to the parity anomaly of the fermion sector \cite{H}. Both these terms are
necessary for complete Poincar\'e invariance of the theory. Rest of the terms contain further
quantum corrections and higher derivatives, which are sub-dominant at the long wavelength limit.
As $V^{\mu\nu}$ contain off-diagonal contributions, there will be contribution from temporal
component $a_0$ to the angular momentum of the spatial ones. Thus the above contains contribution to OAM of
planar photon from both tree level dynamics and vacuum fluctuations.

\section{Schr\"odinger Equation of Photon}
Modification to OAM of the planar gauge field becomes considerably important provided a
first quantized formulation of the system is possible. For a Lorentz covariant vector field, 
such a reduction is not possible in general, owing to the instability of the vacuum ({\it i.e.},
presence of negative norm states). However,
in case of a gauge field, choice of a non-relativistic gauge can reduce its symmetry to a 
Galilean one \cite{IZ}. That is how Maxwell field can couple to a non-relativistic electron.
In both Coulomb and axial gauge, a first quantized equation of motion was obtained 
for planar QED with only the CS term at the tree level \cite{H2}. In such an equation,
modification to OAM, as in Eq. \ref{L6}, alters the centrifugal potential, effecting the
radial dynamics of the `photon'. 
\paragraph*{}Provided suitable gauge-fixing is adopted, an $N$-particle state can be defined 
in the effective gauge sector, stable without introduction of anti-particles, as \cite{H2},

\be
\left\vert N\right\rangle=\int d^2x_1\ldots d^2x_N\phi^\dagger(x_1)\ldots\phi^\dagger(x_N)f(x_1,...,x_N)\vert 0\rangle,\nonumber
\ee
where $f(x_1,...,x_N)$ is the N-particle wave-function. Crucial to this is the existence 
of a unique fermion mass operator, defined as,

\be
M=m\int\psi^\dagger(x)\psi(x)d^2x,\label{02}
\ee
accommodating the mass degeneracy, arising through Bargmann's superselection rule \cite{Bargmann} in a harmless
manner. In effect, $M$ serves as a central extension to the proposed Schr\"odinger algebra \cite{H},
satisfying,

\be
\left[M,\psi(x)\right]=-m\psi(x)\quad\text{and}\quad\left[M,a_i(x)\right]=0=\left[M,a(x)\right].\label{03}
\ee 
For eigenstates $\vert N\rangle$ being the eigenstate of the $N$-particle Hamiltonian,
with eigenvalue (energy) $E$, the corresponding $N$-particle Schr\"odinger equation 
can be written as,

\bea
Ef(x_1,...,x_N)&=&-\frac{1}{2m}\sum_{i=1}^N\left[\nabla_i+ie^2\nabla'_i\sum_{j\neq i}\mathcal{D}(x_{ij})\right]^2f(x_1,...,x_N);\label{SE}\\
{\rm where},~~~x_{ij}&=&x_i-x_j,\nonumber
\eea
in Coulomb gauge \cite{H2}. Here, $\nabla'_i$ acts on the function immediately next to it and
repeated indexes do {\it not} mean summation. The propagator
$\mathcal{D}(x_i-x_j)\equiv\mathcal{D}({\bf x}_i-{\bf x}_j)$ correspond to spatial (${\bf a}(x)$)
or temporal ($a_0(x):=a(x)$) gauge field components, representing dynamics of a
{\it single} particle. The effective Lagrangian in Eq. \ref{L2} now can be resolved in terms of these spatial
and temporal components as,

\begin{align}
\mathcal{L}\equiv&\frac{1}{2}\left\{1+\Pi_e\right\}\left[a\boldsymbol{\nabla}^2a+{\bf a}\cdot\boldsymbol{\nabla}^2{\bf a}+2\dot{a}\boldsymbol{\nabla}\cdot{\bf a}+\left(\boldsymbol{\nabla}\cdot{\bf a}\right)^2+\dot{{\bf a}}\cdot\dot{{\bf a}}\right]\nonumber\\
&+\frac{i}{2}\left\{\Pi_o+i\mu\right\}\left[a\left(\boldsymbol{\nabla}\times{\bf a}\right)+{\bf a}\times\boldsymbol{\nabla} a+{\bf a}\times\dot{{\bf a}}\right];\label{L7}\\
{\rm where},\quad a^{\mu}&=\left(a,a^k\right),~~\dot{x}:=\partial_tx.\nonumber
\end{align}
Application of the non-relativistic Euler-Lagrange equation,

\be
\partial_t\frac{\delta\mathcal{L}}{\delta\dot{\phi}}+\partial_k\frac{\delta\mathcal{L}}{\delta\partial_k\phi}=\frac{\delta\mathcal{L}}{\delta\phi},\nonumber
\ee 
leads to the respective equations of motion for spatial and temporal components as,

\bea
2\ddot{a}^i-\boldsymbol{\nabla}^2a^i-\partial_i\left(\boldsymbol{\nabla}\cdot{\bf a}\right)&=&2\partial_i\dot{a}+i2\left[\frac{\Pi_o+i\mu}{1+\Pi_e}\right]\epsilon^{ij}\partial_ja\label{L8a}\\
\text{and}~~~\boldsymbol{\nabla}^2a&=&2\boldsymbol{\nabla}\cdot\dot{\bf a}-i2\left[\frac{\Pi_o+i\mu}{1+\Pi_e}\right]\boldsymbol{\nabla}\times{\bf a}.\label{L8b}
\eea
The gauge components are interlinked, even in the Coulomb gauge, reflecting the underlying
`actual' Lorentz symmetry, owing to the Maxwell term in the original Lagrangian. This elucidates the fundamental difference 
of the present theory with one with pure CS gauge term \cite{H2}. The fact that the gauge sector is {\it not} merely
topological, dynamics of the effective theory carries the signature of the complete symmetry of the same, the space-time
sectors become interdependent.
The dynamics of respective components are defined by the equation with 
Laplacian of the same, thereby leading to the respective propagators. However, this 
identification excludes the effect of quantum corrections to the respective propagators,
thereby excluding the same from the `photon Schr\"odinger equation'. This is expected from
a non-relativistic, first-quantized approach, with energies far lower than those required
for the excitation of virtual pairs. 

\paragraph*{}More intriguingly, the complete topological sector gets decoupled from the dominant dynamics, including the
tree-level CS contribution, tagged by the coefficient $\mu$. Unlike the case in Ref. \cite{H2}, the gauge sector had a 
tree-level dynamics that overcame the inherently topological CS part that incorporates interaction, as seen in 
the last terms of both the Eq.s \ref{L8a} and \ref{L8b}. As we will see this distinction will prevail as the anomalous 
contribution to OAM of the `effective' photon.

\paragraph*{}The temporal component $a$ satisfies the Laplace's equation, yielding the `static'
propagator $\propto\log\vert {\bf x}_i-{\bf x}_j\vert$ in plane. The presence of dynamics,
however, makes the corresponding result for the spatial counterpart more involved. The 
propagator is time-dependent in general, with the momentum-space representation,

\be
G^{ij}\left(\omega,{\bf k}\right)=\frac{1}{2\omega^2-{\bf k}^2}\left[2\delta^{ij}+\frac{k^ik^j}{\omega^2-{\bf k}^2}\right].\label{N1}
\ee
From Eq. \ref{SE}, as the time-independent Schr\"odinger equation is under consideration, the analysis
is to be confined to the spatial sector with ${\bf a}$ having a fixed energy $\omega$, the latter being
a free parameter. This is {\it not} the single-particle energy, and defines an $N$-particle state
with definite energy, represented by spatial component of the gauge field. This effectively projects
the system onto the position-space, leading to the propagator,  

\be
G^{ij}\left(\omega,{\bf x}-{\bf y}\right):=\int\frac{d^2k}{(2\pi)^2}G^{ij}\left(\omega,{\bf k}\right)\exp\left\{-i{\bf k}.\left({\bf x}-{\bf y}\right)\right\}.\nonumber
\ee
The angular integral identities,

\be
\int_0^{2\pi}\cos\left\{a\cos(\theta)\right\}=2\pi J_0(\vert a\vert) \quad\text{and}\quad \int_0^{2\pi}\sin\left\{a\cos(\theta)\right\}=0,\nonumber
\ee
with $J_n$ being the $n$-th order Bessel function of the first kind, leads to the final expression
of the position space propagator for ${\bf a}$ as,

\begin{align}
G^{ij}\left(\omega,{\bf r}\right)\equiv&i\left[\frac{1}{4\omega r}\left\{\sqrt{2}J_1\left(\sqrt{2}\omega r\right)-J_1(\omega r)\right\}-\frac{1}{2}J_0\left(\sqrt{2}\omega r\right)\right]\delta^{ij}+\frac{1}{4r^2}\left[J_0\left(\sqrt{2}\omega r\right)-J_2\left(\sqrt{2}\omega r\right)\right.\nonumber\\
&-\left.\frac{1}{2}J_0(\omega r)+\frac{1}{2}J_2(\omega r)\right]r^ir^j-\frac{1}{4\omega r^3}\left\{\sqrt{2}J_1\left(\sqrt{2}\omega r\right)-J_1(\omega r)\right\}r^ir^j;\label{N1}\\
{\rm where},&\quad{\bf r}:={\bf x}-{\bf y},~~~r:=\vert{\bf r}\vert.\nonumber
\end{align}
The above expression considerably simplifies to,

\be
G^{ij}\left(\omega,{\bf r}\right)\equiv-\frac{i}{2}J_0\left(\sqrt{2}\omega r\right).\label{L9}
\ee
in the Coulomb (or radiation) gauge, effectively considering the long-range contribution, given 
the present low-energy treatment.

\paragraph*{}The effect of modified OAM can most simply be shown for the two-particle case, with the dynamics given by
Eq. \ref{SE}, which has been reduced to the following form:

\bea
Ef_n\left({\bf x}_1,{\bf x}_2\right)=&&-\frac{1}{2m}\sum_{(i,k)=1}^2\left[\partial_i+ie^2\epsilon^{pl}\sum_{j\neq i}^2\partial^l_iG^{nm}\left({\bf x}_{ij}\right)\right]\nonumber\\
&&\times\left[\partial_k+ie^2\epsilon^{pq}\sum_{r\neq k}^2\partial^q_kG^{ms}\left({\bf x}_{kr}\right)\right]f_s\left({\bf x}_1,{\bf x}_2\right);\label{N02}\\
{\rm with},&&\quad {\bf x}_{ij}={\bf x}_i-{\bf x}_j,\nonumber
\eea
where the repeated indexes $l,~m,~n,~p,~q,~s$ are summed. The notation $\partial^i_j$ denotes differentiation with
respect to the $i$-th component of the $j$-th coordinate, and repeated indexes means summation. Then, usual identities
for Bessel functions,

\bea
\frac{d}{dx}\left[x^mJ_m(x)\right]&=&x^mJ_{m-1}(x),~~~J_{-n}(x)=(-1)^nJ_n(x)\nonumber\\
\text{and}\quad\frac{d}{dx}J_m{x}&=&\frac{1}{2}\left[J_{m-1}(x)-J_{m+1}(x)\right],\nonumber
\eea
finally leads to the two-particle Schr\"odinger equation, 

\begin{align}
Ef_n\left({\bf x}_1,{\bf x}_2\right)\equiv&\frac{1}{2m}\left[{\bf p}^2_1+{\bf p}^2_2+\omega^2e^4\frac{J^2_1\left(\sqrt{2}\omega r\right)}{2r^2}\right.\nonumber\\
&+\left.\frac{\omega e^2}{\sqrt{2}}\left\{\left({\bf x}_1-{\bf x}_2\right)\times\left({\bf p}_1-{\bf p}_2\right),\frac{J_1\left(\sqrt{2}\omega r\right)}{r^2}\right\}\right]f_n\left({\bf x}_1,{\bf x}_2\right),\label{L10}
\end{align}
valid for the spatial gauge component. Here, the curly bracket represents anti-commutator, representing the symmetric
quantum product of two operators.

\begin{figure}
\centering 
\includegraphics[width=3.5 in]{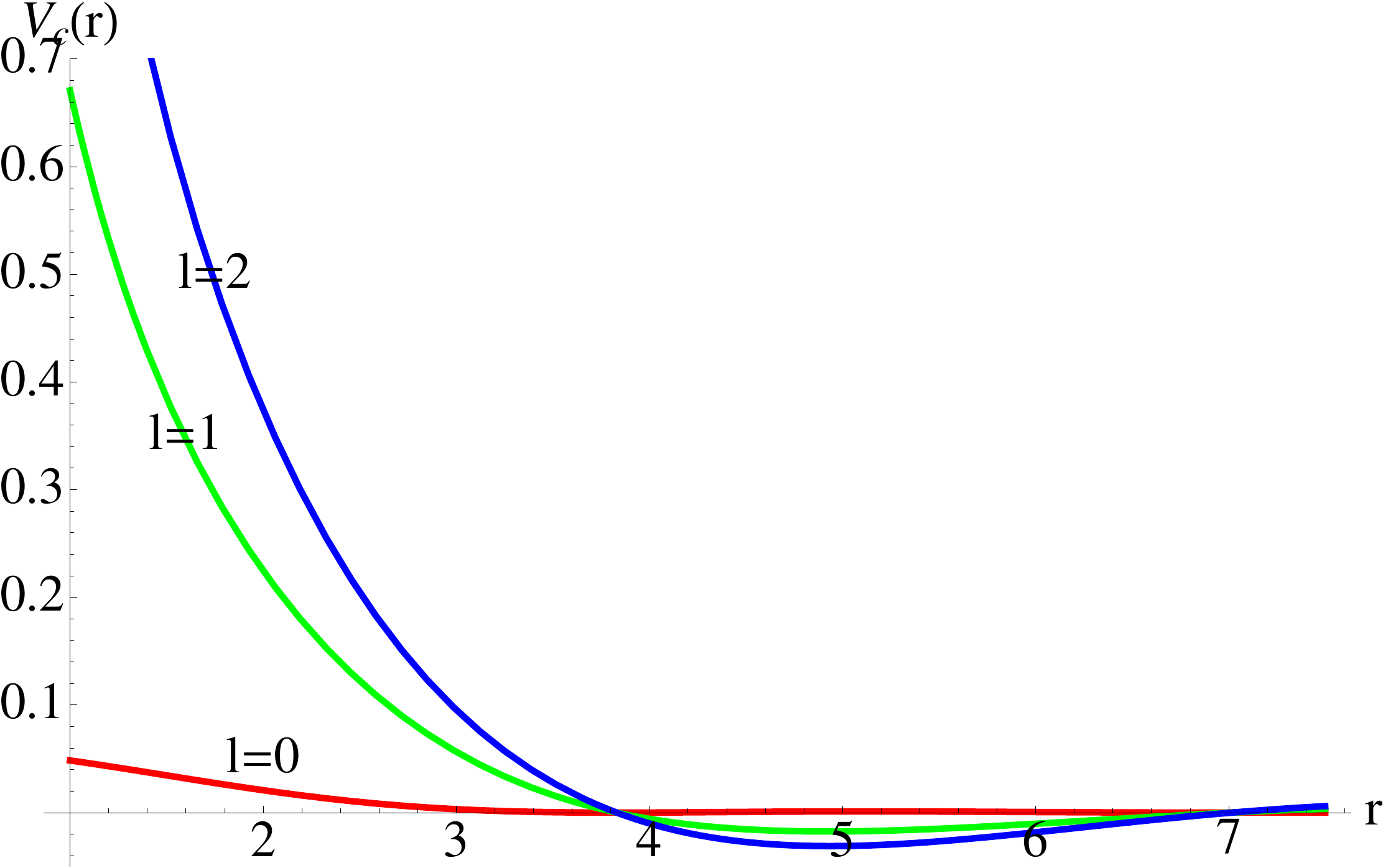}
\caption{Plots of the correction to centrifugal potential $V_c(r)$ for different values of OAM:
$L^2=l(l+1)$, in the ``Schr\"odinger equation'' for photon. This shows possibility of a `centrifugal
resonance'. Here, $\omega=1/\sqrt{2}$ and $e=1$. The presence of the centrifugal repulsion $L^2/r^2$ 
prevents the formation of a shallow bound-state.}\label{f01}
\end{figure}

\begin{figure}
\centering 
\includegraphics[width=4 in]{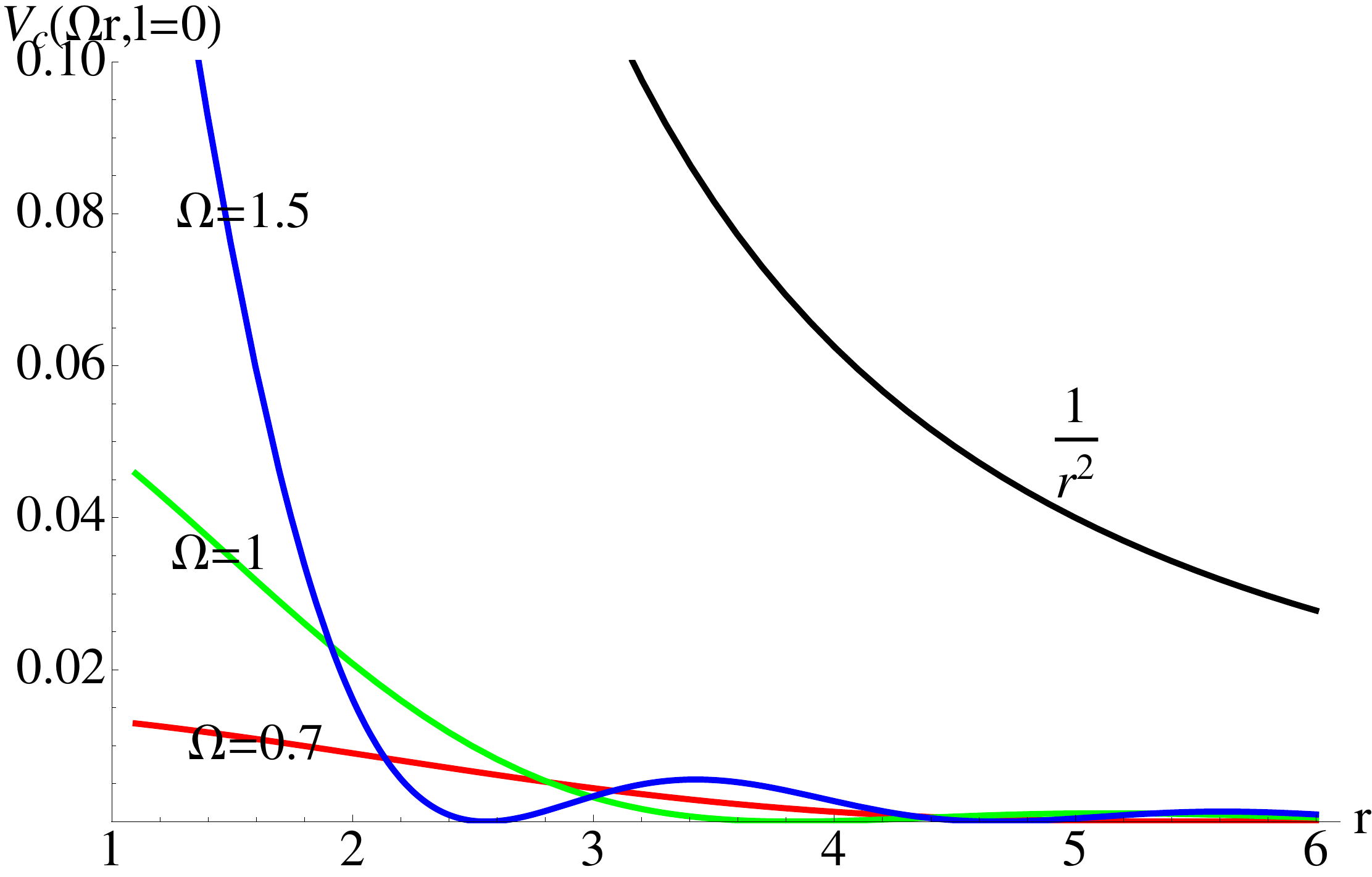}
\caption{Plots of the $s$-wave centrifugal potential, for different values of $\Omega=\sqrt{2}\omega$, and
the inverse-square potential, showing the prior to be of shorter range, required for the existence of
Efimov-like states. Further, suitably high value of $\Omega$ is required to attain a local minimum, leading
to resonance. Here $e=1$.}\label{f02}
\end{figure}

\paragraph*{}An equivalent effective single-particle equation can be obtained by adopting the
center-of-mass (CM) variables \cite{H2}:

\bea
&&{\bf p}_1:=\frac{1}{2}{\bf P}+{\bf p},\quad{\bf p}_2:=\frac{1}{2}{\bf P}-{\bf p};\quad{\bf R}:=\frac{1}{2}\left({\bf x}_1+{\bf x}_2\right),\quad{\bf r}:={\bf x}_1-{\bf x}_2,\nonumber\\
{\rm yielding},&& f_n\left({\bf x}_1,{\bf x}_2\right)=f_n\left({\bf R},{\bf r}\right)\equiv\exp\left(i{\bf P}.{\bf R}\right)\exp\left(iL\theta\right)h^L_n(r),\label{L11}
\eea
leading to the radial Schr\"odinger equation for the present system,

\bea
2m\tilde{E}h^L_n(r)&=&\left[-\frac{1}{r}\frac{d}{dr}r\frac{d}{dr}+\frac{\left\{L+\frac{\omega e^2}{\sqrt{2}}J_1(r)\right\}^2}{r^2}\right]h^L_n(r);\label{L12}\\
\tilde{E}&:=&E-\frac{{\bf P}^2}{4m},~~~J_1(r):=J_1\left(\sqrt{2}\omega r\right),\nonumber
\eea
with ${\bf P}$ being the CM momentum. The angular momentum is shifted by a local function, leading to a non-trivial
centrifugal behavior, interpretable as effective mutual interaction between two {\it effective} spatial photons,
mediated through interaction with low-energy fermions. The spatial photons are the effective description of fermions
interacting via exchange of {\it actual} photons, which can very well be a fermion-anti-fermion bound state at sufficiently
low energies \cite{H,Own2}. Therefore, this anomalous contribution to centrifugal interaction can very well be interpreted
in terms of Coulomb interaction between charged fermions, mediated by photons. This modification to the centrifugal term
is,

\be
V_c(\omega r):=\omega^2e^4\frac{J^2_1\left(\sqrt{2}\omega r\right)}{2r^2}+\sqrt{2}L\omega e^2\frac{J_1\left(\sqrt{2}\omega r\right)}{r^2},\label{L13}
\ee
which is depicted in Fig. \ref{f01}. This `shift' display local minima for certain values of OAM, separated by shallow
barriers. Such regions can capture a system for finite durations, thereby leading to possible
resonances against the centrifugal repulsion $L^2/r^2$, except for in the $l=0$ channel ($s$-wave).
As discussed before, this centrifugal extension is an effect of tree-level dynamics, devoid of topological
contributions. However, this result is exclusive to 2+1 dimensions, as per eq. \ref{L9}. Therefore,
the dynamic and topological sectors separate out in the non-relativistic limit, the latter being 
known to yield constant shift to the OAM \cite{H2} in absence of tree-level dynamics, with a possible
anyonic sector \cite{HagenPRL}.
\paragraph*{Efimov connection:}The local shift to the centrifugal term can be related with the
Efimov physics  \cite{Ef}, the latter being realized for three non-identical particles, with at least
two of the three being almost bound. Then, two particle can experience shallow, near-resonance
bound states with energies obeying constant scaling. The exact nature of the interaction is
immaterial, as long as it is shorter in range than $r^{-2}$ in radial coordinate $r$. In the
present case, the $s$-wave centrifugal potential (Eq. \ref{L13}) satisfies this condition,
as shown in Fig. \ref{f02}, for suitable values of gauge-field energy $\omega$. 

\paragraph*{}The total number of Efimov states up to energy $E\rightarrow0$ can be expressed
as the radial integral,

\be
N_l(E)=\frac{m}{\pi}\int_{r_0}^a\sqrt{E-V_l(r)}dr,~~~\hbar=1.
\ee
with $r_0$ being the range of interaction, $l$ labeling the particular angular momentum channel
and $a$ being the scattering length at ($a\rightarrow\infty$). At very low energies, only the
s-channel ($l=0$) is of importance. The corresponding potential is,

\be
V_l(r)=\frac{1}{2m}\frac{l(l+1)-s_0^2}{r^2},
\ee
leading to $V_0(r)=-s_0^2/2mr^2$, under the WKB approximation. This finally yields the result
$N_0\cong(s_0/\pi)\log(a/r_0)$. Here, $s_0$ is a constant parameter to be tuned suitably. 
The full $s$-wave ($l=0$) centrifugal potential, corresponding to low energies, becomes, 

\be
V_{l=0}(r)=-\frac{1}{2m}\frac{s_0^2}{r^2}+\omega^2e^4\frac{J^2_1\left(\sqrt{2}\omega r\right)}{2r^2}.
\ee
As we are interested in the long-range (asymptotic) behavior, the centrifugal potential 
are sub-dominant to the first term, as can be seen from Fig. \ref{f02}. This leads to the
expression for ground-state degeneracy as,

\be
N_{l=0}(E=0)=\frac{1}{2\pi}\sqrt{\frac{m}{2}}\int_{r_0}^a\frac{dr}{r},
\ee
leading to the exact result for the Efimov case. Therefore, the non-relativistic,
first-quantized, effective treatment of the planar gauge field, interacting with fermions,
leads to Efimov-like {\it resonances}. However, having this interaction
as an inherent three-body effect is not that straight-forward, though an effective theory
may be constructed.

\section{Discussion and Conclusion}
The quantum corrections to OAM, both due to tree-level dynamics of the gauge field and 
vacuum fluctuations, carry the relativistic signature, reflected by the interdependence
of spatial and temporal components. This re-affirms the gauge-dependence of spin part of the total angular momentum
of the gauge field \cite{Greiner}, requiring suitable gauge fixing. More
importantly, quantum fluctuations contribute in a singular way near the 
two-fermion threshold, physically representing the process of two fermions coming close together
and thereby, contributing divergently to the angular momentum of the effective degree 
of freedom. This is in conformity with the excitonic states in planar QED \cite{H,Own2}.
\paragraph*{}The modification to the angular momentum in the effective gauge channel due to tree-level
dynamics is reflected in the first-quantized approximation, though the same is not true
for the contribution from quantum corrections, as expected. This non-relativistic approach
is justified by the adopted Coulomb gauge \cite{H2}, much like the interaction of
electromagnetic field to non-relativistic charged particles. Interestingly, the tree-level 
topological (CS) part separates out and does not contribute to this modified OAM, unlike
the case in Ref. \cite{H2}. The resultant Schr\"odinger equation of the photon displays 
shift in the OAM by a local function, thereby considerably modifying the the centrifugal 
potential at low energies (s-wave), leading to shallow resonances of Efimov type. The
fact that tree-level gauge dynamics alone can effect low-energy resonances may have 
deeper explanations \cite{Pradhan}. 
\paragraph*{}Low-energy, effective planar QED is realizable in condensed matter systems
like graphene \cite{Gra3} and topological insulators \cite{TI}, with emergent gauge interactions \cite{Ando1}.
The emergent Dirac fermions in these systems can allow for direct measurement of the 
modification to OAM, from Eq. \ref{L6}, following the fact that at low energies
($q\rightarrow0$), $\Pi_e\rightarrow0$, and only the topological form factor $\Pi_0$
survives as a constant, representing the Hall conductivity \cite{Neto}. However, the
physical realization of the photon Schr\"odinger equation demands non-relativistic
effective dynamics, and thereby requires different physical systems.


\end{document}